\documentclass[12pt]{article}

\usepackage{amsfonts}
\usepackage{epsfig}
\usepackage{latexsym}
\usepackage{graphicx}
\def\bequ{\begin{equation}}
\def\eequ{\end{equation}}
\def\barr{\begin{array}}
\def\earr{\end{array}}

\def\ben{\begin{equation}}
\def\een{\end{equation}}
\def\bena{\begin{eqnarray}}
\def\eena{\end{eqnarray}}

\def\spa#1{\phantom{\fbox{\rule[-#1cm]{0cm}{0cm}}}}

\def\b1{e^0}

\newcommand{\be}{\begin{equation}}
\newcommand{\ee}{\end{equation}}
\def\bea{\begin{eqnarray}}
\def\eea{\end{eqnarray}}
\def\Tr{\mbox{Tr}}

\def\del {\partial}
\def\nn{\nonumber}

\def\be{\begin{equation}}
\def\ee{\end{equation}}
\def\bea{\begin{eqnarray}}
\def\eea{\end{eqnarray}}

\def\lesssim{\mathrel{\hbox{\rlap{\hbox{\lower4pt\hbox{$\sim$}}}\hbox{$<$}}}}
\def\gtrsim{\mathrel{\hbox{\rlap{\hbox{\lower4pt\hbox{$\sim$}}}\hbox{$>$}}}}

\renewcommand{\thefootnote}{\fnsymbol{footnote}}
\begin{document}

\hfuzz=100pt
\title{{\Large \bf{Fat Magnon}}}
\author{\\Shinji Hirano\footnote{hirano@nbi.dk}
  \spa{0.5} \\
{{\it The Niels Bohr Institute}}
\\ {{\it Blegdamsvej 17, DK-2100 Copenhagen}}
\\ {{\it Denmark}}}
\date{2006}

\maketitle
\centerline{}

\begin{abstract}
We consider a D-brane type state which shares the characteristic of the recently found giant magnon of Hofman and Maldacena. 
More specifically we find a bound state of giant graviton (D3-brane) and giant magnon (F-string), which has exactly the same anomalous dimension as that of the giant magnon. It is described by the D3-brane with electric flux which is topologically a $S^3$ elongated by the electric flux. The angular momentum and energy are infinite, but split sensibly into two parts -- the infinite part precisely the same as that of the giant magnon and the finite part which can be identified as the contribution from the giant graviton.
We discuss that the corresponding dual gauge theory operator is not a simple chain type but rather admixture of the (sub-)determinant and chain types.
\end{abstract}

\renewcommand{\thefootnote}{\arabic{footnote}}
\setcounter{footnote}{0}
%%%%%%%%%%%%%%%%%%%%%%%%%%%%%%%%%%%%%%%%%%
\section{Introduction}
%%%%%%%%%%%%%%%%%%%%%%%%%%%%%%%%%%%%%%%%%%
The recently found giant magnon of Hofman and Maldacena \cite{Hofman:2006xt} supplies us with an important piece of information in the AdS/CFT/spin chain triality \cite{Maldacena:1997re, Gubser:1998bc, Witten:1998qj, Aharony:1999ti} and \cite{Minahan:2002ve, Zarembo:2004hp, Beisert:2004ry}.
The giant magnon is the string theory dual of magnon in the infinite  spin chain at large 't Hooft coupling. 

The spin chain concerns the states in the string theory on $AdS_5\times S^5$ which carry angular momenta in $S^5$. The chain length is set by the angular momentum $J$ of one's choice in $S^5$. 
So the corresponding states in ${\cal N}=4$ super Yang-Mills (SYM) contain $Z^{J'\le J}$ where $Z$ is a complex adjoint scalar, and the $J-J'$ insertions of other complex scalars correspond to the spin excitations.
The spin chain Hamiltonian corresponds to the dilatation operator in ${\cal N}=4$ SYM. The range of spin interactions correlates with the order of the perturbation in 't Hooft coupling $\lambda$. At the one-loop order the spin interaction is the nearest neighbor and at two-loop the next to the nearest neighbor, and so on. For instance, for the smallest closed set of states ($SU(2)$ subsector), the spin chain is the XXX$_{1/2}$ Heisenberg chain at one-loop  \cite{Minahan:2002ve} and Inozemtsev  chain \cite{Inozemtsev:1989yq} at two-loop \cite{Serban:2004jf}. 
The magnons are elementary excitations in the spin chain and carry the momentum $p$.
They are also a convenient set of states in order to diagonalize the spin chain Hamiltonian, and constitute the essential basis for the Bethe ansatz.

In fact the chain can be identified as the spatial extension of the string in the gauge in which the angular momentum $J$ is uniformly distributed over the string. The magnon is then an excitation of the string, and the momentum $p$ is the worldsheet momentum of the excited string.\footnote{The total momentum of physical excitations must vanish due to the translation invariance on the worldsheet. So the single magnon with nonvanishing $p$ is not physical, corresponding to the fact that the gauge theory operator representing the magnon is ${\cal O}_p\sim\sum_l e^{ipl}\left(\cdots ZZWZZ\cdots\right)$ which is not traced thus not gauge-invariant. We will come back to this point later. }
The $\alpha'$ of the string in $AdS_5\times S^5$ is proportional to $1/\sqrt{\lambda}$. So naively the $\alpha'$ expansion corresponds to the strong 't Hooft coupling expansion. However, the string states of our interest carry the angular momentum $J$. 
In the ingenious limits such as BMN/pp-wave limit \cite{Berenstein:2002jq} (including near BMN corrections of \cite{Callan:2003xr})\footnote{Indeed the three-loop discrepancy was encountered and left unresolved.} and multi-spin strings of \cite{Frolov:2003qc} where the ratio $\lambda'=\lambda/J^2$ is fixed as $\lambda$ and $J$ taken to infinity, the energy of the semi-classical excited string turns out to admit the $\lambda'$ expansion, rendering it possible to compare the string theory and gauge/spin chain results in the weak 't Hooft coupling expansion.
The existence of such double scaling limits is quite remarkable. However, it would be necessary to go beyond these limits in order to understand the integrability of the full string theory on 
$AdS_5\times S^5$.
In particular the magnon has been poorly understood in the string theory side, except for the low momentum case. As the magnon is an essential element in the Bethe ansatz, it would be an important step to understand the magnons in the string theory side. 

At large 't Hooft coupling the spin chain is very long-ranged. The spin chain Hamiltonian/dilatation operator is practically incalculable, as it requires the all-loop SYM computation.  
Nevertheless, for the infinite chain ($J\to\infty$), the all-loop (asymptotic) Bethe ansatz was guessed 
 from the spin chain perspective guided by the integrability, BMN scaling, and a few loop order results in SYM  \cite{Beisert:2004hm, Beisert:2005fw}.  
Remarkably the all-loop guess was later derived by Beisert only by the use of supersymmetry without need of knowing the detailed dynamics of ${\cal N}=4$ SYM \cite{Beisert:2005tm} except for the inspiring inputs from it.\footnote{There is another very intriguing development in this direction \cite{Rej:2005qt}. 
The Hubbard model confirms the all-loop guess  \cite{Beisert:2004hm} of Beisert, Dippel, and Staudacher (BDS), up to the order at which the wrapping interaction would invalidate it. This model is particularly interesting, for it is short-ranged and capable of dealing with the finite chain. At more conceptual level, the Hubbard model might be suggesting a more convenient set of degrees of freedom to describe the theory.} 
This is a significant result. In particular the (asymptotic) S-matrix was determined almost completely up to a phase factor. Incidentally fixing this phase factor is one of the current major issues. However, we will not discuss about it in this paper.

This nonpertubative gauge theory result makes it possible to compare the string theory and gauge/spin chain results far from the BMN type limits.
The limit taken here is instead $N\to\infty$ and then $J\to\infty$ with $\lambda$ kept finite.
In particular, the dispersion relation of magnons shows the distinct momentum dependence -- the energy is periodic in momentum:  
\be
E-J=\sqrt{1+{\lambda\over\pi^2}\sin^2{p\over 2}}\ ,
\label{dispersion_relation}
\ee
for a single magnon.

At first sight it might appear that the periodicity in momentum suggests the discrete worldsheet in which the lattice spacing is to set the period.
However, as it turned out, the magnon is dual to a macroscopic open string orbiting in $S^5$ (the giant magnon), and the momentum $p$ is the geometric angle between two endpoints of the string, 
in accordance with the fact that $p$ is canonical conjugate to the angular momentum $J$  \cite{Hofman:2006xt}. The periodicity of $p$ then follows without discretizing the worldsheet.
The upshot is that the giant magnon precisely reproduces the large 't Hooft coupling limit of the dispersion relation/anomalous dimension (\ref{dispersion_relation}):
\be
E-J={\sqrt{\lambda}\over\pi}\left|\sin{p\over 2}\right|\ ,
\ee
provided that $p$ is not too small.

We might then hope that further studies of the giant magnons will lead us to the better understanding of the integrability and Bethe ansatz for the $AdS_5\times S^5$ string.

Several works on the giant magnons have appeared. One of the directions under study is the multi-magnon bound states initiated by Dorey \cite{Dorey:2006dq} and followed by \cite{Minahan:2006bd, Spradlin:2006wk, Bobev:2006fg, Kruczenski:2006pk, Okamura:2006zv}. The bound states correspond to the giant magnons  with two or three angular momenta in $S^5$. The reference \cite{Minahan:2006bd} also discussed the dual giant magnon which stretches in $AdS_5$ corresponding to the magnon in the $SL(2)$ sector. An important direction to pursue is to calculate the stringy corrections. This was done at one-loop in the limiting case of $J_1\ll J_2$ for the giant magnon with two angular momenta, finding the agreement that it is absent \cite{Minahan:2006bd}.
More challenging but conceptually important question is to understand the case of finite chain. This question was studied in an interesting work \cite{Arutyunov:2006gs}. 
Partly related to this work, the classical (closed) strings with finite $J$ were further studied in \cite{Okamura:2006zv}.
The generalization to a deformed background ($\beta$-deformation) was made in \cite{Chu:2006ae, Bobev:2006fg}, and also the M-theory generalization was studied in \cite{Bozhiloov:2006bi}.

In this paper we consider a D-brane type state which shares the characteristic of the giant magnon.
The D-brane type states play import roles in AdS/CFT. A well-known example is the giant graviton which is a spherical D3-brane rotating in $S^5$ and expanded either in $S^5$ \cite{McGreevy:2000cw} or $AdS_5$ \cite{Grisaru:2000zn}. The giant gravitons are degenerate states with a graviton state propagating in $S^5$. In the semi-classical approximation, at low energy $E\sim{\cal O}(1)$ the adequate description is provided by the graviton, while at high energy $E\sim{\cal O}(N)$ by the giant gravitons . 
In the dual CFT the former corresponds to the trace operator $\Tr Z^J$ and the latter to the multi-trace operators $\Tr_A Z^J$ and $\Tr_S Z^J$, where the subscripts $A$ and $S$ denote the antisymmetric and symmetric representations respectively. 

Another example is the D-branes corresponding to the Wilson loop operators -- the D3-brane of $AdS_2\times S^2$ shape \cite{Drukker:2005kx} and the D5-brane of $AdS_2\times S^4$ shape \cite{Yamaguchi:2006tq}. They are the bound states of a D-brane and fundamental strings.
The former corresponds to the Wilson loop in the symmetric representation $\Tr_S U$ and the latter in the antisymmetric representation $\Tr_A U$, where $U=P\exp\left(\int_C ds(A_{\mu}\dot{x}^{\mu}+\Phi|\dot{x}|)\right)$ \cite{Gomis:2006sb}.\footnote{The D-brane description is valid when the number $k$ of fundamental strings is large. In particular, in the case of the D3 \lq\lq giant" loop, the more correct statement is that it is dual to the Wilson loop $\Tr_{S^k} U$ in the $k$-th symmetric representation which at large $k$ is indistinguishable from  the multiply wound Wilson loop $\Tr U^k$.}

The D-brane type state for the giant magnon we will discuss in this paper can be thought of as a bound state of giant magnons (F-strings) and a giant graviton (D3-brane) expanded in $S^5$. So clearly it is closely related to the giant graviton, but at the same time being the bound state of a D3-brane and F-strings, it is also akin to the \lq\lq giant" Wilson loop.\footnote{There is another example of bound states of giant gravitons. In the presence of the NS-NS $B$-field, the giant in the plane-wave background can form a bound state with D1-branes. The D1-branes wrap on the $S^3$ and squash the sphere giant \cite{Prokushkin:2004pv}.}

The organization of the paper is as follows. In section 2 we find the D-brane type state for the giant magnon as a classical solution in the low energy effective theory of the D3-brane in the $AdS_5\times S^5$ background. We call this object the fat magnon. The anomalous dimension, $E-J$, is shown to be the same as that of the giant magnon. We provide evidence for the interpretation that the fat magnon is a threshold bound state of the giant graviton and giant magnons in the limit we are taking.
We then discuss the dual CFT operator to the fat magnon. Our proposal is not complete. We outline the main ingredients to construct this operator and discuss a possibility.
In section 3 we briefly conclude our results.

%%%%%%%%%%%%%%%%%%%%%%%%%%%%%%%%%%%%%%%%%%%%%%%%%%
\paragraph{Note Added:} The fat magnon in the plane-wave background was previously found by Sadri and Sheikh-Jabbari and called giant hedge-hog \cite{Sadri:2003mx}.

%%%%%%%%%%%%%%%%%%%%%%%%%%%%%%%%%%%%%%%%%%%%%%%%%%%
\section{Fat magnon}
%%%%%%%%%%%%%%%%%%%%%%%%%%%%%%%%%%%%%%%%%%%%%%%%%%%

We wish to find a D-brane type state which shares the characteristic of the giant magnon.
The D-brane suitable for this purpose will be the (topologically) spherical D3-brane of the giant graviton type, since we consider the state with angular mometum $J$ in $S^5$.
To be a magnon-like state, it is essential to have the characteristic geometric angle in $S^5$ for the object of our concern, which corresponds to the magnon momentum $p$. So the D3-brane may lie in $S^5$ rather than $AdS_5$.  However, the giant graviton does not have an open angle. To develop such an angle, the spherical D3-brane ought to be elongated. The stretch of this deformation will be parameterized by the geometric angle. This can be done by turning on an  electromagnetic flux on the brane. In this case it would be natural to turn on an electric flux, thus attaching a F-string to the D3-brane, since the giant magnon is a F-string. 
Then the D-brane state so constructed will be the bound state of the giant graviton and giant magnon, which naturally inherits the property of the giant magnon.
We anticipate that its anomalous dimension $E-J$ will be exactly the same as that of the giant magnon, since $E-J$ is zero for the giant graviton and the only contribution would come from the flux, provided that this is a marginal BPS bound state. This is indeed the case, as we will see below.

We call this bound state the fat magnon. The fatness is that of $S^2$. As is familiar, despite being topologically spherical, it is stable due to the Myers effect \cite{Myers:1999ps} in the presence of the RR five-form field strength. 
%%%%%%%%%%%%%%%%%%%%%%%%%%%%%%%%%%%%%%%%%
\subsection{The string theory side -- probe analysis}
%%%%%%%%%%%%%%%%%%%%%%%%%%%%%%%%%%%%%%%%%

We work in the probe approximation. Since we are interested in the finite size probes with infinite energy, the approximation is, strictly speaking, not justified. Nevertheless, we would still expect the quantitative accuracy of the computation, as is often the case for the BPS configurations.

%%%%%%%%%%%%%%%%%%%%%%%%%%%%%%%%%%%%%%
\subsubsection{Giant magnon}

\begin{figure}[ht]
\begin{center}
$\begin{array}{c@{\hspace{1in}}c}
\multicolumn{1}{l}{\mbox{ }} &
        \multicolumn{1}{l}{\mbox{ }} \\ [-0.6cm]
\epsfxsize=3in
\epsffile{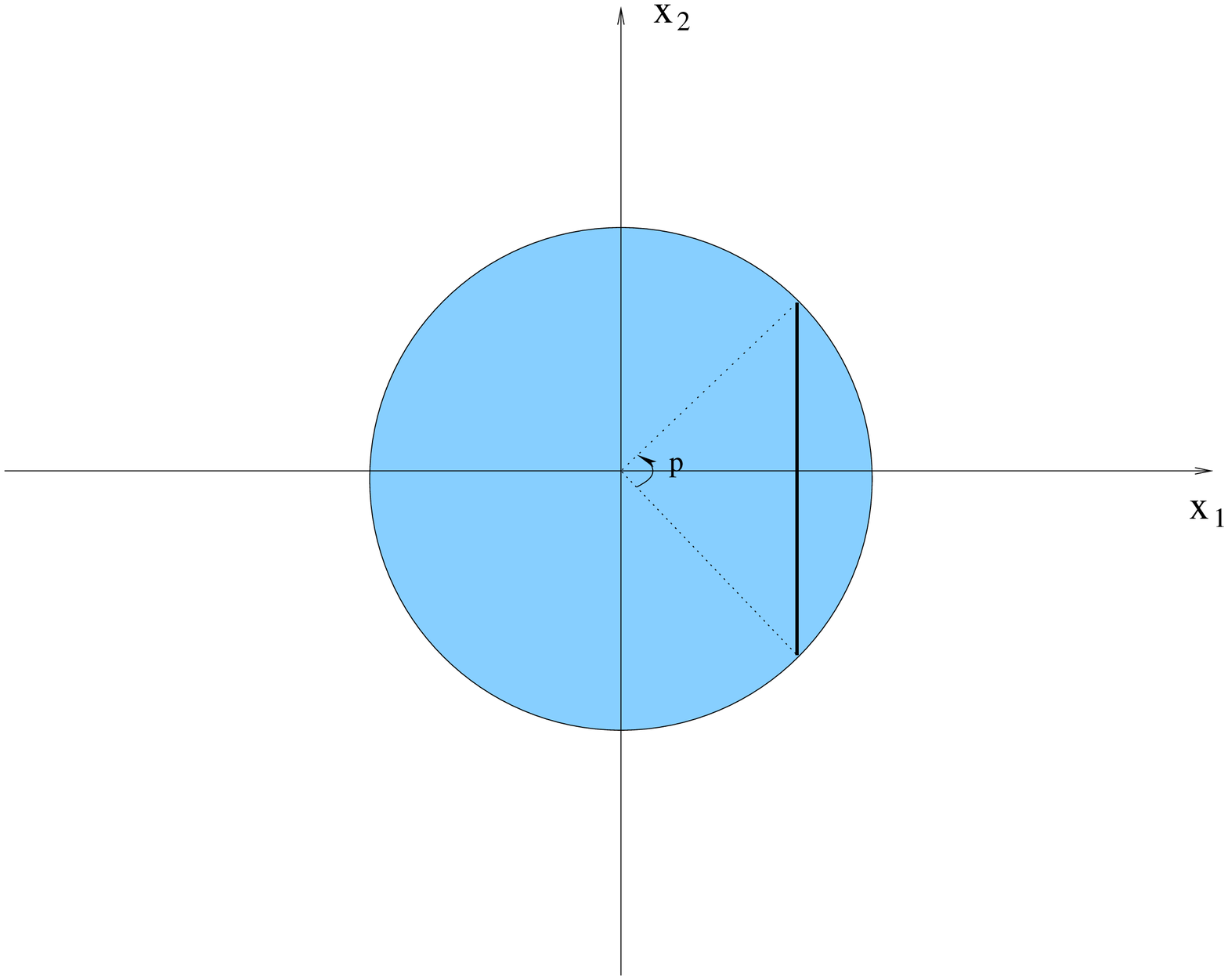} &
        \epsfxsize=2in
        \epsffile{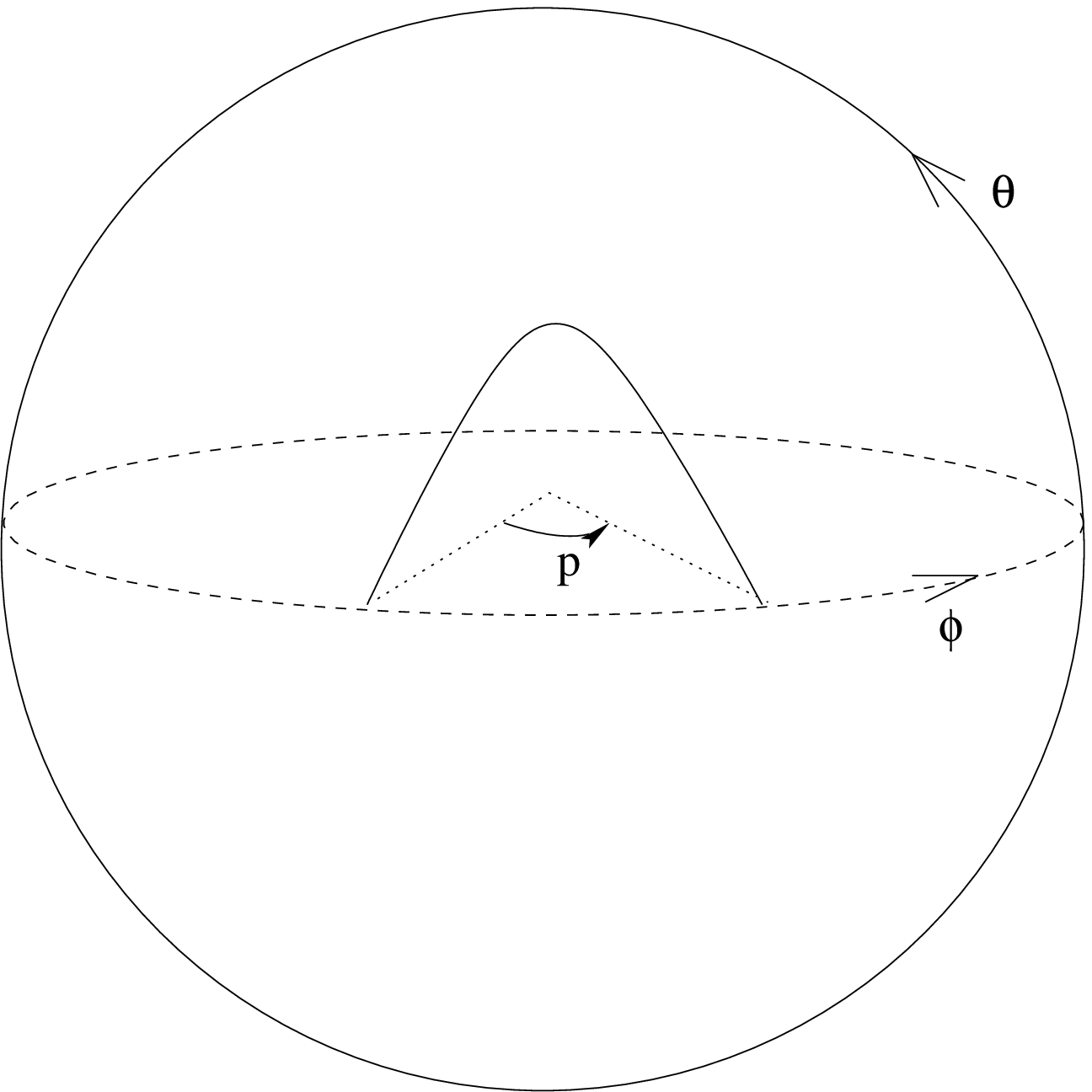} \\ [0.1cm]
\mbox{\bf (A)} & \mbox{\bf (B)}
\end{array}$
\end{center}
\caption{(A) The giant magnon in LLM coordinates: The metric (\ref{LLMmetric}) corresponds to the inside of the disk (droplet). The thick straight line represents a giant magnon. (B) The giant magnon in the standard spherical coordinates.}  
\label{giantmagnon}
\end{figure}

We begin with a brief review of the giant magnon \cite{Hofman:2006xt}. The giant magnon is a macroscopic open string orbiting in $S^5$ and whose endpoints sit on the equator, as shown in Figure \ref{giantmagnon} (B).
We adopt the coordinate system by Lin, Lunin, and Maldacena (LLM) \cite{Lin:2004nb} which turns out to be particularly convenient for our purpose.

The relevant part of the spacetime is $\mathbb{R}\times S^5$. In the LLM coordinates, it reads 
\be
ds^2\Biggr|_{\rho=0}=R^2\left[-\left(1-r^2\right)
\left(dt-{r^2\over 1-r^2}d\widetilde{\phi}\right)^2
+{dr^2+r^2d\widetilde{\phi}^2\over 1-r^2}
+\left(1-r^2\right)\left(d\chi^2+\sin^2\chi 
d\widetilde{\Omega}_2^2\right) 
\right]\ ,
\label{LLMmetric}
\ee
where $\rho=0$ indicates that we are focusing on the geometry at the center of the global $AdS_5$, that is, $\mathbb{R}\times S^5$. $R$ denotes the radius of $S^5$ and $AdS_5$. If we set $r=\cos\theta$ and $\widetilde{\phi}=\phi-t$, we recover the standard coordinate system for $\mathbb{R}\times S^5$.

We denote the worldsheet coordinates by $(\tau, \sigma)$ and choose the static gauge $t=\tau$. We now make the ansatz for the shape and dynamics of the string as
\be
r=r(\sigma)\ ,\qquad
\phi=\phi(\tau,\sigma)\ ,
\ee
where $\phi=\widetilde{\phi}+t$. 

Then the Nambu-Goto action yields
\be
S_{NG}=-{\sqrt{\lambda}\over 2\pi}\int d\tau d\sigma
\sqrt{{r'^2\over 1-r^2}+r^2\phi'^2-{r'^2r^2\over 1-r^2}\dot{\phi}^2}\ ,
\ee
where we have used the fact that $R^4=4\pi g_sNl_s^4=\lambda l_s^4$. The dash and dot denote the derivative with respect to $\sigma$ and $\tau$ respectively.

Let us introduce the Cartesian coordinates 
$(x_1,x_2)=(r\cos\widetilde{\phi},r\sin\widetilde{\phi})$, in terms of which we have
\bea
r'={x_1x_1'+x_2x_2'\over r}\ ,\qquad
\phi'={x_1x_2'-x_2x_1'\over r^2}\ .\nn
\eea
We wish to find the solution with $\dot{\phi}=1$. We further make the ansatz that $x_1(\sigma)$ is constant. 
Indeed the equations of motion with respect to $\phi$ and $r$ are automatically fulfilled. So $x_2(\sigma)$ is an arbitrary function. We now fix the residual gauge freedom $\sigma\to\widetilde{\sigma}(\sigma)$ by choosing $x_2(\sigma)=a\sigma+b$ where $a$ and $b$ are constants.
We then impose the boundary condition that both ends of the string reach the edge ($r=1$) of the droplet. Let us also fix the range of $\sigma$ to be $0\le\sigma\le\pi$. Then the solution yields
\be
x_2(\sigma)={2\over\pi}\sqrt{1-x_1^2}\left(\sigma-{\pi\over 2}\right)\ ,\qquad
x_1=\mbox{const}\ .
\label{gmsol}
\ee
Since the action is invariant under the translation of $\phi$ (or $\widetilde{\phi}$), any pair $(x_1,x_2)$ obtained from (\ref{gmsol}) by a rotation is a solution.

One can readily see that the angular momentum $J=\int d\sigma \pi_{\phi}$ 
where $\pi_{\phi}=\del{\cal L}/\del\dot{\phi}$ as well as the energy 
$E=\int d\sigma (\dot{\phi}\pi_{\phi}-{\cal L})$ diverge. 
For later use, we give the explicit formula for the angular momentum,
\be
\pi_{\phi}={\sqrt{\lambda}\over 2\pi}{x_2^2x_2'\over 1-r^2}\ .\label{gmangularmom}
\ee
However, their difference $E-J$ remains finite and yields the magnon dispersion relation at large 't Hooft coupling,
\be
E-J=-\int d\sigma{\cal L}={\sqrt{\lambda}\over \pi}\sqrt{1-x_1^2}
={\sqrt{\lambda}\over \pi}\sin{p\over 2}\ ,\label{gmanomalousD}
\ee
where $p$ is the geometric angle between two endpoints of the string, as sketched in Figure \ref{giantmagnon}. 

%%%%%%%%%%%%%%%%%%%%%%%%%%%%%%%%%%%%%%%%%%%%%
\subsubsection{Fat magnon}

\begin{figure}[ht!]
\centering \epsfysize=3cm
\includegraphics[scale=0.4]{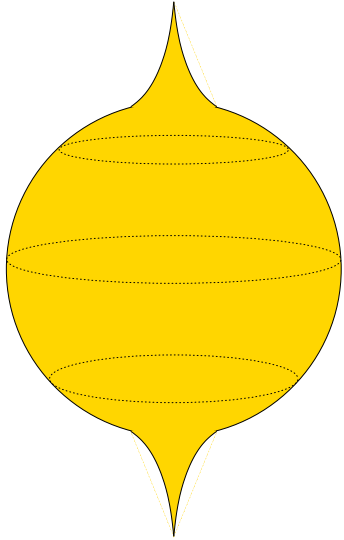}
\caption{A sketch of fat magnon} 
\label{fmsketch}
\end{figure}

As motivated above, we consider a topologically spherical D3-brane with electric flux. 
So what we will find is a variant of BIon of \cite{Callan:1997kz}. Two endpoints of the giant magnon bound to the giant graviton corresponds to a pair of unlike electric charges put at the antipodal points in $S^3$. They will develop the spikes as in the BIon case. So the fat magnon will look like the Figure \ref{fmsketch}.\footnote{The giant hedge-hog in \cite{Sadri:2003mx} shares the same properties.}

Let the worldvolume coordinates be $(\tau,\sigma_1,\sigma_2,\sigma_3)$.
We work in the static gauge $t=\tau$ and the embedding of the D3-brane into (\ref{LLMmetric}) as
\be
\chi=\sigma_1\equiv \sigma\ ,\quad
\widetilde{\Omega}_2=(\sigma_2,\sigma_3)\ .
\ee
We make the ansatz for the shape and dynamics of the D3-brane as
\be
r=r(\sigma)\ ,\qquad 
\phi=\phi(\tau,\sigma)\ .
\ee
This is the same form as in the case of the giant magnon.
Note that this ansatz assumes the $SO(3)$ symmetry of $S^2$.

With this ansatz the D3-brane action yields
\bea
S_{D3}&=&-T_3\left[\int d\tau d^3\sigma\sqrt{-\det(G_{\mu\nu}\del_aX^{\mu}\del_bX^{\nu}
+2\pi l_s^2F_{ab})} +\int C_4\right]\nn\\
&=&-4\pi R^4T_3\int d\tau d\sigma\sin^2\sigma
\left[(1-r^2)\sqrt{{\cal D}}-(1-r^2)^2\dot{\phi}\right]\ ,
\eea
where 
\be
{\cal D}={r'^2\over 1-r^2}+r^2\phi'^2-{r'^2\over 1-r^2}r^2\dot{\phi}^2
-\left({2\pi l_s^2\over R^2}\right)^2F_{\tau\sigma}^2+(1-r^2)(1-r^2\dot{\phi}^2)\ ,
\label{calD}
\ee
and $T_p={2\pi\over g_s(2\pi l_s)^{p+1}}$. 
So the effective tension is $T\equiv 4\pi R^4T_3={2\over\pi}N$, and $2\pi l_s^2/R^2=2\pi/\sqrt{\lambda}$. 

We further make the following ansatz to find a solution:
\be
x_1=\mbox{const}\ ,\qquad
{2\pi \over\sqrt{\lambda}}F_{\tau\sigma}=\pm x_2'\ .
\label{ansatz}
\ee
Again we wish to find the solution with $\dot{\phi}=1$. Then the equation of motion with respect to $\phi$ yields
\be
{d\over d\sigma}\left(\sin^2\sigma\, x_2'\right)=0\ .\label{x2eom}
\ee
The equation of motion with respect to $r$ is then trivially satisfied, and the $A_{\sigma}$ equation of motion yields
\be
{d\over d\sigma}\left(\sin^2\sigma\, F_{\tau\sigma}\right)=0\ ,
\ee
which is equivalent to (\ref{x2eom}), given the above ansatz.

Thus the ansatz is consistent and the solution is given by
\be
x_2=c-\kappa\cot\sigma\ ,\label{solution}
\label{x2}
\ee
where $c$ and $\kappa$ are the constants. As we will see shortly, the constant $\kappa$ is fixed by the flux quantization. Although $\kappa$ can be either positive or negative, we will consider the positive case for definiteness unless otherwise stated.

\begin{figure}[ht!]
\centering \epsfysize=3cm
\includegraphics[scale=0.4]{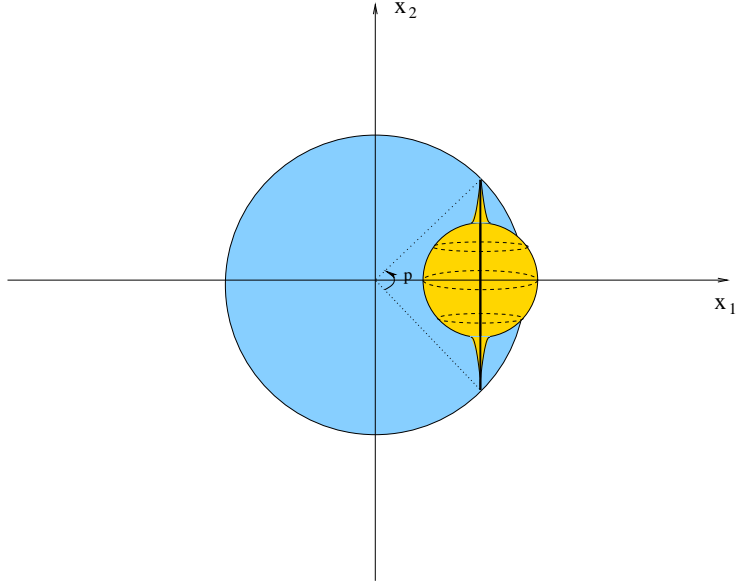}
\caption{A sketch of fat magnon in LLM coordinates} 
\label{fatmagLLM}
\end{figure}

As is clear from the metric (\ref{LLMmetric}), the radius squared of $S^3$ is $R^2(1-r^2)=R^2(1-x_1^2-x_2^2)$. Thus the solution (\ref{solution}) implies that the $S^3$ is elongated along the $\chi=\sigma$ direction. 
Let the range of $\sigma$ be $\sigma_0\le\sigma\le\pi-\sigma_0$. In accordance with the giant magnon, we impose the boundary condition that the fat magnon stretches all the way between the edge of the droplet, that is, $x_2(\sigma_0)=-\sqrt{1-x_1^2}$ and $x_2(\pi-\sigma_0)=\sqrt{1-x_1^2}$.
So in LLM coordinates the fat magnon looks like the Figure \ref{fatmagLLM}. The thickness of the fat magnon grows in the directions orthogonal to the $(x_1,x_2)$ plane.

Now the electric flux must be quantized. The quantization yields
\be
\pi_{A}\equiv{\del{\cal L}\over\del\dot{A_{\sigma}}}=\pm{4N\over\sqrt{\lambda}}\kappa
=k\ ,
\ee
where $k$ is an integer and the number of F-strings. Hence the constant $\kappa$ is determined as
\be
\kappa=\pm{\sqrt{\lambda}\over 4N}k\ .
\ee

As in the case of the giant magnon, the energy and angular momentum diverge. 
In fact the momentum density is given by
\be
\pi_{\phi}\equiv{\del{\cal L}\over\del\dot{\phi}}=
T\sin^2\sigma\left[{r^2r'^2\over 1-r^2}+(1-r^2)\right]\ ,\label{angular}
\ee
and the Hamiltonian (density) by 
${\cal H}=F_{\tau\sigma}\pi_A+\dot{\phi}\pi_{\phi}-{\cal L}=kF_{\tau\sigma}+\pi_{\phi}$.
However, their difference is finite and given by
\be
E-J=\int d\sigma F_{\tau\sigma}\pi_A
={kR^2\over{2\pi l_s^2}}\int_{-\sqrt{1-x_1^2}}^{\sqrt{1-x_1^2}} 
dx_2
=k{\sqrt{\lambda}\over \pi}\sin{p\over 2}\ .
\label{anodimfat}
\ee
This is $k$ multiple of the anomalous dimension of the giant magnon (\ref{gmanomalousD}). 
Indeed since $k$ is the number of F-strings,\footnote{The strings are uniformly smeared over $S^2$.} the single string case precisely agrees with the giant magnon result, as expected.
The $k>1$ case corresponds to the superposition of $k$ giant magnons with the same momentum $p$.

As the calculation tells, the only contribution to the anomalous dimension comes from the electric flux, thus from the F-strings/giant magnons. The contribution from the D3-brane is absent.  As mentioned above, $E-J$ is zero for the giant graviton. This may suggest that two contributions are simply additive. 
That would be the case if the fat magnon is a marginal bound state of the giant graviton and giant magnon. We will now provide further evidence for this observation.

%%%%%%%%%%%%%%%%%%%%%%%%%%%%%%%%%%%%%%%%%%%
\subsubsection{A closer look at the fat magnon}

The formula (\ref{angular}) for the angular momentum contains more information than just being singular. We will see that it is composed of two parts -- the part precisely the same as the giant magnon angular momentum and the contribution from the giant graviton. 

The first term diverges at the edge $r=1$ of the droplet, but it is precisely the same as the giant magnon angular momentum density (for $k=1$). 
To see it, we rewrite the first term as
\be
\pi_{\phi}^{gm\subset fm}\equiv T\sin^2\sigma{r^2r'^2\over 1-r^2}
=T\sin^2\sigma{x_2^2x_2'^2\over 1-r^2}
=\pm k{\sqrt{\lambda}\over 2\pi}{x_2^2x_2'\over 1-r^2}\ .
\label{gminfmmom}
\ee
This is indeed $k$ multiple of the giant magnon angular momentum density (\ref{gmangularmom}).

We now argue that the second term $\pi^{gg\subset fm}_{\phi}$ of $\pi_{\phi}$ is the contribution from the giant graviton. It can be evaluated as
\be
q\equiv\int_{\sigma_0}^{\pi-\sigma_0} 
d\sigma\pi^{gg\subset fm}_{\phi}=N\left(1-{2\over\pi}\sigma_0\right)\left(
\sin^2{p\over 2}-\kappa^2\right)+{N\over\pi}\sin(2\sigma_0)\left(
\sin^2{p\over 2}+\kappa^2\right)\ ,
\ee
where $\sigma_0=\tan^{-1}\left({\kappa\over\sin(p/2)}\right)$ setting $c=0$ in (\ref{x2})
and $0\le\sigma_0\le{\pi\over 2}$. Here we have implicitly restricted to the case $\kappa>0$, or equivalently the plus sign with positive $k$ in (\ref{gminfmmom}). This is reflected in the orientation of the fat magnon we have chosen. The other choice would have yielded the minus of this result.

%The length $J$ of the chain should not too much exceed $N$, since the corresponding gauge theory operators would then have broken into multi-traces.
%In fact the limit was taken first $N\to\infty$ and then $J\to\infty$ so that we can conform to this condition.
%This in particular implies that $k\sqrt{\lambda}\ll N$, since otherwise $J(=\int d\sigma \pi_{\phi}^{gm\subset fm})$ would have been much larger than $N$, as we can see it from (\ref{gminfmmom}). 
%Thus the number $k$ of F-strings should not be too large and we must impose $\kappa\ll 1$.
%In fact, as we are in the $N\to\infty$ limit, we would also have to take the limit $\kappa\to 0$.

Recall that we are in the strict $N\to\infty$ limit. The constant $\kappa$ is then taken to zero, provided that $k\sim{\cal O}(1)$.
In order to comply with our boundary condition, this limit must be taken keeping 
$\kappa\cot\sigma_0=\sin{p\over2}$ fixed. To summarize we take the limit
\be
\kappa\ ,\sigma_0\to 0\qquad\mbox{keeping}\quad\kappa\cot\sigma_0
=\sin{p\over2}\quad\mbox{fixed}\ .\label{limit}
\ee
In this limit the giant graviton angular momentum becomes
\be
q\to N\sin^2{p\over 2}\ .\label{limitggmom}
\ee
The angular momentum $q$ must be quantized and an integer, but the ratio $q/N$ can take a continuous value in the large $N$ and $q$ limit (classical limit).

\begin{figure}[ht!]
\centering \epsfysize=3cm
\includegraphics[scale=0.4]{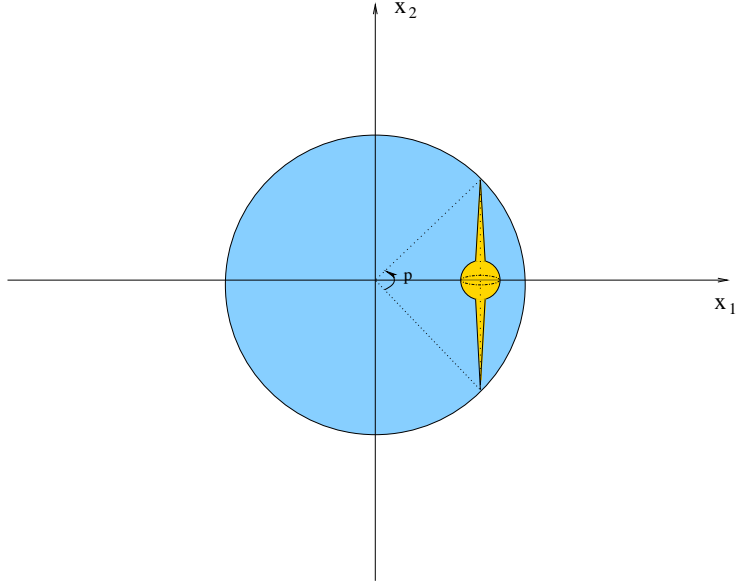}
\caption{A sketch of fat magnon in LLM coordinates near the limit (\ref{limit})} 
\label{fatmagLLMlimit}
\end{figure}

Recall that the size $R_{gg}$ of the giant graviton is related to its angular momentum $q$ by 
$R_{gg}=R\sqrt{{q\over N}}$ \cite{McGreevy:2000cw, Grisaru:2000zn}. So in the current case, the size of the giant would be $R_{gg}=R\sin{p\over 2}$. In fact, as we can see from (\ref{x2}) with $c=0$, $x_2$ is almost zero except at the ends of the range of $\sigma$ in our limit. This means that the fat magnon is almost a perfect $S^3$ but it develops sharp spikes at the north and south poles, as depicted approximately in Figure \ref{fatmagLLMlimit}. Indeed the size of $S^3$ away from the poles is $R\sqrt{1-r^2}= R\sin{p\over 2}$, in accordance with the relation between the size and angular momentum of the giant graviton.

We have not checked the supersymmetries of the fat magnon explicitly. However, it is quite likely that the fat magnon is BPS in the limit (\ref{limit}). A more detailed look at the D3-brane action provides supports for it. The DBI part of the D3-brane action has a square root factor. Typically the inside of the square root becomes a perfect square for BPS solutions. This is indeed the case for the fat magnon: ${\cal D}$ defined in (\ref{calD}) becomes $(1-r^2)^2$. Moreover, the Lagrangian density ${\cal L}$ is vanishing. The same happens for the BPS giants. In that case, the vanishing Lagrangian results in the BPS saturation $E=J$. 
In the fat magnon case, as we have seen, the energy is equal to $E=J_{gg}+J_{gm}+\Delta$, where $J_{gg}$ and $J_{gm}$ are the angular momenta of the giant graviton and giant magnon respectively, and $\Delta$ is the anomalous dimension (\ref{anodimfat}). So the energy is a simple sum of the giant graviton energy $E_{gg}=J_{gg}$ and giant magnon energy $E_{gm}=J_{gm}+\Delta$. Each one of them is BPS. So we may conclude that the fat magnon is a threshold BPS bound state.

Finally we comment on the validity of our approximation. 
The curvature of the fat magnon behaves near the spikes ($\sigma\sim 0, \pi$) as
\be
{\cal R}_{fm}\sim {1\over R^2 \sin^2\sigma (1-r^2)}\stackrel{\kappa\to 0}
{\longrightarrow}{1\over R^2\sin^2{p\over 2}\sin^2\sigma} \ .
\ee
Thus the curvature becomes large. This also implies that the energy density becomes large. So strictly speaking, the probe approximation is not valid near the spikes. The probe approximation breaks down also in the giant magnon case. The energy density diverges at the endpoints in that case.
In the fat magnon case, however, the description in terms of the DBI$+$CS action also breaks down near the spikes. The $\sigma$ derivatives of the collective coordinate $x_2$ and the field strength $F_{\tau\sigma}$ blow up at the spikes. So generally speaking, there will be higher derivative corrections to be taken into account.
However, we believe that, as is often the case for BPS configurations, those corrections are protected from being generated and the probe and DBI$+$CS approximation can still provide the accurate results.

%%%%%%%%%%%%%%%%%%%%%%%%%%%%%%%%%%%%%%%%%%%
\subsection{The gauge theory side -- dual CFT operator}
%%%%%%%%%%%%%%%%%%%%%%%%%%%%%%%%%%%%%%%%%%%

We wish to find a conceivable proposal for the dual CFT operator for the fat magnon. 
There are three elements in the character of the fat magnon which would compose the basis for a possible proposal: (1) giant magnon, (2) sphere giant (giant gravion in $S^5$), (3) attaching the giant magnon (open string) to the giant graviton (D3-brane).
So the logical step to take is to understand the dual CFT operators for (1) the giant magnon, (2) sphere giant, and (3) sphere giant with open strings attached, and combine them together.
Indeed each one of them is known:

\medskip\noindent
(1) The CFT operator dual to the giant magnon takes the Bethe ansatz form \cite{Minahan:2002ve, Zarembo:2004hp, Beisert:2004ry}, 
\be
\left({\cal O}_p\right)_{j}^{\mbox{ }i}=\sum_l e^{ipl}\left(\cdots ZZWZZ\cdots\right)_{j}^{\mbox{ }i}\ ,
\label{gmoperator}
\ee
where $l$ denotes the location of $W$. 
Note that it is not traced. If it was, the phase factor $e^{ipl}$ would have been trivial.  
In other words, we do not impose the cyclic invariance on the spin chain states, in order to have nonzero single magnon momentum. Two indices left uncontracted mark the endpoints of the giant magnon, a macroscopic open string.

\medskip\noindent
(2) The operator dual to the sphere giant is a (sub-)determinant operator (or the trace over an antisymmetric representation) \cite{Balasubramanian:2001nh, Corley:2001zk}.
\be
{\cal O}^{gg}=\epsilon^{i_1i_2\cdots i_{q-1} i_{q}}_{j_1j_2
\cdots j_{q-1} j_{q}}Z_{i_1}^{\mbox{ }j_1}Z_{i_2}^{\mbox{ }j_2}\cdots Z_{i_q}^{\mbox{ }j_q}\ ,
\label{sphgiantop}
\ee
where $\epsilon^{i_1i_2\cdots i_{q-1} i_{q}}_{j_1j_2
\cdots j_{q-1} j_{q}}\equiv q!\delta_{[j_1}^{[i_1}\delta_{j_2}^{i_2}\cdots\delta_{j_{q-1}}^{i_{q-1}}\delta_{j_{q}]}^{i_{q}]}=\sum_{\sigma\in S_{q}}(-1)^{\sigma}\delta_{\sigma(j_1)}^{i_1}\delta_{\sigma(j_2)}^{i_2}\cdots\delta_{\sigma(j_{q-1})}^{i_{q-1}}\delta_{\sigma(j_{q})}^{i_{q}}$, 
and the square bracket denotes the anti-symmetrization. Incidentally this operator can be rewritten in terms of multi-trace operators.

The reason for this operator being the dual of sphere giant may be understood by mapping the giant gravitons into the matrix ($Z$) quantum mechanics/free fermion system \cite{Corley:2001zk, Berenstein:2004kk, Lin:2004nb, Jevicki:1991yi}. 
The eigenstates of the matrix quantum mechanics Hamiltonian are given by the characters $\chi_{R}(Z)$ for the representation $R$. 
In terms of the free fermions, the sphere giant is a hole, while the AdS giant (giant graviton in $AdS_5$) is a particle. The former translates to the antisymmetric, and the latter the symmetric representation. The length of column and row in the Young tableaux corresponds to the energy/angular momentum of the giant. 
Incidentally the former being the antisymmetric representation results in the existence of the maximal angular momentum/energy/size of the sphere giant, that is, $q_{max}=N$ -- stringy exclusion principle \cite{McGreevy:2000cw}. This accords with the fact that the energy of a hole is bounded from above, set by the Fermi energy.   

\medskip\noindent
(3) The operator dual to the sphere giant with an open string excitation was conjectured to be \cite{Balasubramanian:2002sa}
\be
{\cal O}^{gg+open}=\epsilon^{i_1i_2\cdots i_{q-1} i_{q}}_{j_1j_2
\cdots j_{q-1} j_{q}}Z_{i_1}^{\mbox{ }j_1}Z_{i_2}^{\mbox{ }j_2}\cdots 
Z_{i_{q-1}}^{\mbox{ }j_{q-1}}{\cal O}[\Phi_i, D_lZ]_{i_q}^{\mbox{ }j_q}\ ,
\label{giantwopen}
\ee
where $D_l$ is the covariant derivative in $\mathbb{R}\times S^3$, and ${\cal O}[\Phi_i, D_lZ]$ is a monomial (\lq\lq word") composed of the real adjoint scalars $\Phi_{i=1,\cdots, 6}$ and $D_lZ$, corresponding to the open string excitation. 

This is based on the observation that the frequencies of small fluctuation modes on the giant do not depend on its size, that is, $\omega_k=(k,k+1,k+2)/R$ where $k$ is the angular momentum in $S^3$ the worldvolume of the giant \cite{Das:2000st}. There are three patterns depending on in which direction the giant vibrates. This peculiar property of the vibration modes ensures that the simple insertion of the operator of the type ${\cal O}[\Phi_i, D_lZ]$ yields the right quantum numbers.

However, there is a restriction on the form of ${\cal O}[\Phi_i, D_lZ]$. The beginning and end of the word cannot be $Z$. By applying the expansion 
\bea
\epsilon^{i_1i_2\cdots i_p}_{j_1j_2\cdots j_p}
=\sum_{q=1}^{p}(-1)^{p-q}\delta_{j_q}^{i_p}
\epsilon^{i_1i_2\cdots i_{q-1}i_qi_{q+1}\cdots i_{p-1}}_{j_1j_2\cdots j_{q-1}j_{q+1}\cdots\cdot\cdot j_p}\ ,\nn
\eea
it is straightforward to show that
\bea
&&(p-1)\epsilon^{i_1i_2\cdots i_{q-1}i_{q+1}\cdots i_{p-1}i_q}_{j_1j_2\cdots j_{q-1}j_{q+1}
\cdots j_{p-1} j_{p}}
Z_{i_1}^{j_1}\cdots Z_{i_{q-1}}^{j_{q-1}}Z_{i_{q+1}}^{j_{q+1}}\cdots Z_{i_{p-1}}^{j_{p-1}}
(Z{\cal O}')_{i_q}^{j_p}\nn\\
&=&\epsilon^{i_1i_2\cdots i_{p-1}}_{j_1j_2\cdots  j_{p-1}}
Z_{i_1}^{j_1}\cdots  Z_{i_{p-1}}^{j_{p-1}}\Tr({\cal O}')-
\epsilon^{i_1i_2\cdots i_p}_{j_1j_2\cdots j_p}Z_{i_1}^{j_1}Z_{i_2}^{j_2}\cdots Z_{i_{p-1}}^{j_{p-1}}
({\cal O}')_{i_p}^{j_p}\ ,
\label{subdetformula}
\eea
where $Z{\cal O}'={\cal O}$, and a similar formula holds for the case of $Z$ at the end.
This means that if $Z$ sits at the beginning or end of the word, the operator (\ref{giantwopen}) breaks into the sphere giant ($\epsilon Z\cdots Z$) with a closed string emission ($\Tr{\cal O}'$) and a larger giant with an open string excitation ($\epsilon Z\cdots Z{\cal O}'$) on it. 
So in that case the operator (\ref{giantwopen}) is not an independent operator.

\medskip

We are now in a position to make a proposal. The most naive guess for the operator dual to the fat magnon would be\footnote{This type of operators was previously considered in \cite{Agarwal:2006gc}.}
\be
{\cal O}^{fat}_p\stackrel{??}{=}\lim_{\stackrel{N\to\infty, q\to\infty}{q/N=\sin^2{p\over 2}}}
\epsilon^{i_1i_2\cdots i_{q} i_{q+1}}_{j_1j_2
\cdots j_{q} j_{q+1}}Z_{i_1}^{\mbox{ }j_1}
Z_{i_2}^{\mbox{ }j_2}\cdots Z_{i_{q}}^{\mbox{ }j_{q}}
\left({\cal O}_p\right)_{i_{q+1}}^{\mbox{ }j_{q+1}}\ ,
\label{fmoperatorwrong}
\ee
where ${\cal O}_p$ is the dual CFT operator for the giant magnon (\ref{gmoperator}).
Note that the giant graviton momentum $q$ must be equal to $N\sin^2{p\over 2}$, as explained in (\ref{limitggmom}).

There appear to be two problems in this proposal; (1) The beginning and/or end of the word ${\cal O}_p$ are/is $Z$. The repeated use of the above formula yields the sum of many giants plus closed string emission (and a single maximal giant with $W$). (2) This operator can be rewritten in terms of the (multi-)traces. In this case it implies that the phase factor $e^{ipl}$ is trivial. Either way the anomalous dimension of this operator cannot depend on $p$.

We need to find a way to evade these problems. 
Given the fact that the gauge theory operator dual to the giant magnon is non-gauge invariant (see (\ref{gmoperator})), 
we might as well consider the non-gauge invariant operator for the fat magnon. A possibility we propose is
\be
\left({\cal O}^{fat}_p\right)_{j_{0}}^{\mbox{ }i_{0}}
\stackrel{?}{=}\lim_{\stackrel{N\to\infty, q\to\infty}{q/N=\sin^2(p/2)}}
\epsilon^{i_{0}i_1i_2\cdots i_q i_{q+1} }_{ j_{0}j_1j_2
\cdots j_q j_{q+1}}Z_{i_1}^{\mbox{ }j_1}
Z_{i_2}^{\mbox{ }j_2}\cdots Z_{i_q}^{\mbox{ }j_q}
\left({\cal O}_p\right)_{i_{q+1}}^{\mbox{ }j_{q+1}}\ .
\label{fmoperator2}
\ee
In this case the phase factor $e^{ipl}$ in ${\cal O}_p$ does not yield trivial, evading the second of the problems faced above.   
A formula similar to (\ref{subdetformula}) still holds in this non-gauge invariant case.
However, the repeated application of the formula would not lead us to the linear dependence, if at all, of this operator in any obvious way, due to the non-triviality of the summation over $l$ with the phase factor $e^{ipl}$. So it seems to evade the first of the above problems too.

%
%\be
%{\cal O}^{fat}_p
%=\lim_{\stackrel{N\to\infty, q\to\infty}{q/N=\sin^2{p\over 2}}}
%\epsilon^{i_1i_2\cdots i_q i_{q+1}}_{j_1j_2
%\cdots j_q j_{q+1}}Z_{i_1}^{\mbox{ }j_1}
%Z_{i_2}^{\mbox{ }j_2}\cdots Z_{i_q}^{\mbox{ }j_q}
%\left(\bar{Z}{\cal O}_p\bar{Z}\right)_{i_{q+1}}^{\mbox{ }j_{q+1}}\ .
%\label{fmoperator2}
%\ee
%

The corresponding proposal for the case of $k>1$ would then be 
\be
\left({\cal O}^{fat}_{kp}\right)_{j_{0}}^{\mbox{ }i_{0}}
\stackrel{?}{=}\lim_{\stackrel{N\to\infty, q\to\infty}{q/N=\sin^2(p/2)}}
\epsilon^{i_{0}i_1i_2\cdots i_q i_{s_1}\cdots i_{s_k} }_{ j_{0}j_1j_2
\cdots j_q j_{s_1}\cdots j_{s_k}}Z_{i_1}^{\mbox{ }j_1}
Z_{i_2}^{\mbox{ }j_2}\cdots Z_{i_q}^{\mbox{ }j_q}
\left({\cal O}_p\right)_{i_{s_1}}^{\mbox{ }j_{s_1}}
\cdots \left({\cal O}_p\right)_{i_{s_k}}^{\mbox{ }j_{s_k}}\ .
\label{fmoperator2multi}
\ee

In order to construct physical objects of the fat magnon type, we need to combine multiple of them together, connecting one end after another to eventually close the loop, In the case of giant magnons,  
the corresponding physical CFT operator is an appropriate superposition (determined by the Bethe Ansatz) of the following type of operators:
\be
\sum_{l_1,\cdots, l_k} e^{i\left(p_1l_1+\cdots +p_kl_k\right)}\Tr\left(\cdots ZZWZ\cdots ZWZZ\cdots\right)\ ,
\ee
where $p_1+\cdots +p_k=0$, and there are $k$ insertions of $W$s at the locations $l_1,\cdots,l_k$.
This is equivalent to 
\be
\left({\cal O}_{p_1}\right)_{i_{s_1}}^{\mbox{ }j_{s_1}}\left({\cal O}_{p_2}\right)_{j_{s_1}}^{\mbox{ }i_{s_2}}
\cdots \left({\cal O}_{p_{k-1}}\right)_{i_{s_{k-2}}}^{\mbox{ }j_{s_{k-1}}}
\left({\cal O}_{p_k}\right)_{j_{s_{k-1}}}^{\mbox{ }i_{s_{1}}}
\ee
by allowing the locations $l_i$ to be anywhere in the whole chain beyond the $i$-th chain.

So it seems natural to propose that the physical CFT operator for the fat magnons be an appropriate superposition of the operators
\be
\left({\cal O}^{fat}_{p_1}\right)_{i_{s_1}}^{\mbox{ }j_{s_1}}\left({\cal O}^{fat}_{p_2}\right)_{j_{s_1}}^{\mbox{ }i_{s_2}}
\cdots \left({\cal O}^{fat}_{p_{k-1}}\right)_{i_{s_{k-2}}}^{\mbox{ }j_{s_{k-1}}}
\left({\cal O}^{fat}_{p_k}\right)_{j_{s_{k-1}}}^{\mbox{ }i_{s_{1}}}
\ee
with $p_1+\cdots +p_k=0$ and allowing the locations $l_i$ in ${\cal O}_{p_i}$ (in ${\cal O}^{fat}_{p_i}$) to be anywhere in the longer chain whenever several ${\cal O}_{p_i}$s connect.

However, at this stage we may state that it is currently not well-understood how to precisely combine multiple fat magnons to build a physical object.

%The credibility of this proposal remains to be seen.

%%%%%%%%%%%%%%%%%%%%%%%%%%%%%%%%%%%%%%%%%%%%%%%%%%%
\section{Conclusion}
%%%%%%%%%%%%%%%%%%%%%%%%%%%%%%%%%%%%%%%%%%%%%%%%%%%
We found a new D-brane type state in AdS/CFT/spin chain triality. It is a bound state of the giant graviton (D3-brane) and giant magnons (F-strings), and has the same anomalous dimension as that of the giant magnons. 
In other words, the giant magnons can become fat by the Myers effect due to the 5-form RR flux. It is also a generalization of BIon in a curved background ($S^5$) carrying the angular momentum.

There are a few obvious directions to pursue. 
It would be interesting to consider the generalization to the bound state of multi-magnons, {\it i.e.}, the giant magnon with two or three angular momenta.
On this score, an explicit check of the supersymmetry ($\kappa$-symmetry) of the fat magnon is preferable, and it would help us to find the generalization to the multi magnon bound states.
Also it would be nice to understand the scattering of fat magnons.

We discussed a possible form of the dual CFT operator for the fat magnon. Given that the fat magnon is a bound state of the giant graviton and giant magnon, it is quite conceivable that the CFT operator is an admixture of the (sub-)determinant and chain type. Although our proposal is incomplete, there does not seem to be much room for the operator to take the form other than (\ref{fmoperator2}). Yet clearly the further study is required.

More importantly it is desirable to understand the relevance of the fat magnon to the spin chain system. This question might require us to study the length varying spin chain of  \cite{Berenstein:2005vf,Agarwal:2006gc,Berenstein:2005fa}.

%%%%%%%%%%%%%%%%%%%%%%%%%%%%%%%%%%%%%%%%%%%%%%%%%%
\section*{Acknowledgment}
The author would like to thank Sumit Das, Nadav Drukker, Troels Harmark, Joe Minahan, Shin Nakamura, Vasilis Niarchos, and Simon Ross for useful discussions and comments. He is also grateful to University of Porto, especially Carlos Herdeiro, for their warm hospitality where this project was initiated.

%%%%%%%%%%%%%%%%%%%%%%%%%%%%%%%%%%%%%%%%%%%%%%%%%%%%%%%%%%%%%%%%%%%

\end{document}